\documentclass[twocolumn,superscriptaddress,longbibliography,aps,
pra,preprintnumbers,10pt]{revtex4-2}

\usepackage[normalem]{ulem}
\usepackage{graphicx}
\usepackage{bm}
\usepackage{color}
\usepackage{epstopdf}
\usepackage{amsmath}
\usepackage{amssymb}
\usepackage{epstopdf}

\usepackage{multirow}

\usepackage[urlcolor=blue,colorlinks=true,citecolor=blue,linkcolor=blue,pdfstartview={FitH},bookmarks=false]{hyperref}

\usepackage{xcolor}
\graphicspath{{fig/}{./fig/}{.}}

\begin{document}

\title{Majorana bound states in the presence of the half-smeared potential}

\author{Surajit Basak}
\email[e-mail: ]{surajit.basak@ifj.edu.pl}
\affiliation{\mbox{Institute of Nuclear Physics, Polish Academy of Sciences, W. E. Radzikowskiego 152, PL-31342 Krak\'{o}w, Poland}}

\author{Andrzej Ptok}
\email[e-mail: ]{aptok@mmj.pl}
\affiliation{\mbox{Institute of Nuclear Physics, Polish Academy of Sciences, W. E. Radzikowskiego 152, PL-31342 Krak\'{o}w, Poland}}

\date{\today}

\begin{abstract}
The Majorana bound state can be realized in one dimensional chain, in form of two well localized and separated states at both ends of the chain.
In this paper, we discuss the case when the potential is smeared at one end of the system.
In our investigation, we assume the smearing in form of a quadratic function of position.
We show that the smearing potential lead to the emergence of extra in-gap states, and effectively decrease the local gap (around the smeared potential).
The Majorana states are still preserved in the system, however, their localization depend on the smearing.
Moreover, the symmetric localization of the Majorana states from both side of the system is no longer preserved in the presence of the smearing potential. 
\end{abstract}

\maketitle

\section{Introduction}

The idea of realization of the Majorana quasiparticles at the ends of a chain, introduced by Alexei Kitaev~\cite{kitaev.01}, started a period of intensive studies of this issue~\cite{leijnse.flensberg.12,bena.17,aguado.17,lutchyn.bakkers.18,pawlak.hoffman.19}.
Several setups to realize the Majorana bound states (MBS) were recently explored -- e.g. monoatomic magnetic chain deposited on the superconducting surface~\cite{nadjperge.drozdov.14,pawlak.kisiel.16,ruby.heinrich.17,jeon.xie.17,feldman.randeria.17,steinbrecher.rausch.18,kamlapure.cornils.18,kim.palaciomorales.18,schneider.brinker.20} or hybrid superconductor/semiconductor devices~\cite{mourik.zuo.12,das.ronen.12,deng.yu.12,albrecht.higginbotham.16,deng.vaitiekenas.16,gul.zhang.17,chen.yu.17,chen.woods.19}.

Realization of the MBS is an attractive topic due to its predicted application in constructing topologically protected Majorana-based qubits~\cite{aasen.hell.16}.
Decoherence-free quantum computing operation is related to the ability of storing information nonlocaly~\cite{nayak.simon.08}.
In the standard picture, this protection is guaranteed by a high degree of spatial nonlocality of the MBS.
Although localized in space themselves, the majorana quasiparticle, together with its pair, behaves like a single fermion, nonlocal in space.
By a higly nonlocal MBS, we mean that the MBS at the ends of the wire have almost zero overlap with each other.
The degree of Majorana nonlocality $\eta^{2}$ can be understood as a quantity denoting overlapping of the MBS wavefunctions~\cite{penaranda.aguado.18}.
Highly nonlocal MBS are characterized by $\eta \rightarrow 0$ (absence of overlapping), while $\eta \rightarrow 1$ (maximal overlapping) indicates that MBS are not so well localized at the boundaries.
On the other hand, overlapping of the wavefunction is related to the observed MBS energy.
Experimentally, the Majorana nonlocality can be measured from splitting due to hybridization of the zero modes in resonance with a quantum dot state at one end of the nanowire~\cite{deng.vaitiekenas.18}.

Now, we shortly describe the typical hybrid superconductor/semiconductor device studied experimentally.
The device is based on semiconducting nanowires with epitaxial superconductor layer on three facets of the wire, grown by molecular beam epitaxy~\cite{krogstrup.ziino.15}.
Such a prepared system is characterized by the hard induced superconducting gap~\cite{chang.albrecht.15}.
Electrostatic control of wire and barrier density is provided by side gates and a global back gate.
A quantum dot (QD) can also be realized in this set up at the bare end (not covered by superconductor) of the semiconducting wire~\cite{chen.woods.19}.
Occupancy of the QD is tuned by the voltage on the gates close to the dot region.
In a natural way, the spatial profile of the gate voltage lead to a non-homogeneous carrier distribution~\cite{kobialka.ptok.19}.
In this paper, we will discuss, how smearing of the potential at the one end of nanowire affect the MBS properties.

Modification of the potential along the system can be expected in several situations.
For example, in presence of the gate potentials (mentioned in the previous paragraph) or in the case of topological superfuild in the optical trap.
In the first case, the attachment of the wire to a two-dimensional plaquette allows the emergence of a pair of zero-energy edge states (one localized at the end of the wire and the second one localized at the edge of plaquette)~\cite{kobialka.domanski.19}.
However, experimental realization of such system indicate formation of the zero-mode from coalescing Andreev bound states~\cite{suominen.kjaergaard.17}.
In the second case, the shape of the optical trap potential strongly affect the realized topological phase~\cite{liu.hu.12,liu.drummond.12,xu.mao.14,liu.15,ptok.cichy.18}.
Summarizing, the local potential in the system has a huge impact on the MBS properties~\cite{fleckenstein.dominguez.18,moore.zeng.18,cao.zhang.19}, and can lead to a modification of the MBS localization.
Finally, it is worth mentioning that the inhomogeneity introduced due to the local potential~\cite{rainis.trifunovic.13,liu.sau.17,huang.pan.18,moor.stansecu.18,stanescu.tewari.19,pan.dassarma.20}, or attachment of the QD to the nanowire~\cite{ptok.kobialka.17,reeg.dmytruk.18,kobialka.ptok.19} can lead to the emergence of additional in-gap states.

\begin{figure}[!b]
\includegraphics[width=\linewidth]{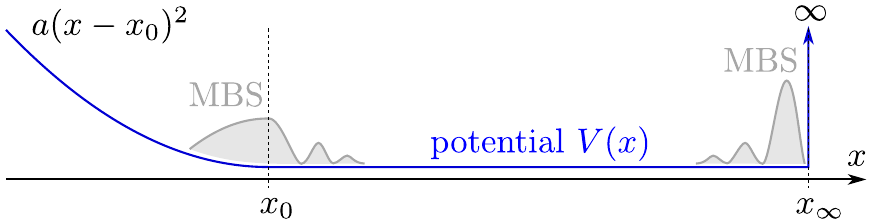}
\caption{
Schematic representation of the discussed system with half-smeared potential $V(x)$ (blue line). 
The Majorana bound states (MBS) are induces at the end of the system.
Localization of the MBS strongly depends on the shape of $V(x)$.
\label{fig.schem}
}
\end{figure}

In context of the argument mentioned above, it is fair to ask how the smearing of the local potential at one end of nanowire affect the MBS properties.
This paper is organized as follows.
First, we briefly describe the used model and technique (Sec.~\ref{sec.theo}).
Next, in Sec.~\ref{sec.num} we present and discuss our numerical results which are concluded in Sec.~\ref{sec.sum}.

\section{Theoretical background}
\label{sec.theo}

In this paper, we discuss the half-smeared potential presented in Fig.~\ref{fig.schem}.
From right side of the nanowire, we assume sharp step of the potential $V(x) \rightarrow \infty$, while from left side parabolic-like potential $V(x) \sim ( x - x_{0} )^{2}$.
In central part part of the system we assume homogeneous potential:
\begin{eqnarray}
\label{eq.pot} V(x) = \left\lbrace \begin{array}{cl}
a ( x - x_{0} )^{2} , & \text{for} \quad x < x_{0} , \\ 
0 , & \text{for} \quad x_{0} \leq x \leq x_{\infty} , \\ 
+\infty , & \text{for} \quad x_{\infty} < x ,
\end{array} \right.
\end{eqnarray}
Here, $a$ is a free parameter to control the smearing.

We describe our system by the following Hamiltonian:
\begin{eqnarray}
\mathcal{H} &=& \mathcal{H}_{0}  + \mathcal{H}_{SO}  + \mathcal{H}_{SC} + \mathcal{H}_{V} .
\end{eqnarray}
Here, $\mathcal{H}_{0} = \sum_{i,j\sigma} \left( - t \delta_{\langle i,j \rangle} - \left( \mu + \sigma h \right) \delta_{ij} \right) c_{i\sigma}^{\dagger} c_{j\sigma}$
denotes the kinetic term, which describes the hopping  of the electrons between the nearest-neighbor sites.
$c_{i\sigma}$ ($c_{i\sigma}^{\dagger}$) denotes the anihilation (creation) operator of the electron with spin $\sigma$ at site $i$.
Here $t$ is the hopping integral, $\mu$ is chemical potential, while $h$ is a Zeeman magnetic filed.
We neglect orbital (diamagnetic) pair-breaking effects~\cite{kiczek.ptok.17}.
The spin-orbit coupling is descrbied by $\mathcal{H}_{SO} = - i \lambda \sum_{ i \sigma \sigma' } c_{i\sigma} \left( \hat{\sigma}_{y} \right)_{\sigma\sigma'} c_{i+1\sigma'} + h.c.$, where $\hat{\sigma}_{y}$ is the Pauli {\it y}-matrix.
The superconductivity is described by the BCS-like term $\mathcal{H}_{SC} = \sum_{i} \left( \Delta c_{i\downarrow} c_{i\uparrow} + h.c. \right)$, where $\Delta$ denotes superconducting gap. 
Finally, we introduce the half-smeared potential $\mathcal{H}_{V} = \sum_{i\sigma} V ( i ) c_{i\sigma}^{\dagger} c_{i\sigma}$, where we assume $V(x)$ in the form~(\ref{eq.pot}), where $x \equiv | {\bm R}_{i} |$ denotes position of the site $i$. 
This system is presented schematically in Fig.~\ref{fig.schem}.
Here, the free parameter $a$ can be treated as a parameter controlling the smearing of the potential $V(x)$. 
For $1/a \rightarrow 0$, we get exactly 1D chain, while for $a \rightarrow 0$, we have half-open chain.

In the presence of $V(x)$, the system is highly non-homogeneous.
Thus, the quasiparticle spectrum of $\mathcal{H}$ can be obtained from a diagonalization procedure based on the Bogoliubov--Valatin transformation:
\begin{eqnarray}
\label{eq.transformacja}
c_{i\sigma} = \sum_{n} \left( u_{in\sigma} \gamma_{n} - \sigma v_{in\sigma}^{\ast} \gamma_{n}^{\dagger} \right) ,
\end{eqnarray}
where $\gamma_{n}$ and  $\gamma_{n}^{\dagger}$ are the ``new'' quasiparticle fermionic operators.
The coefficients $u_{in\sigma}$ and $v_{in\sigma}$ satisfy the Bogoliubov--de Gennes (BdG) equations $\mathcal{E}_{n} \Psi_{in} = \sum_{j} \mathbb{H}_{ij} \Psi_{jn}$~\cite{degennes.89}, where $\Psi_{in} = \left( u_{in\uparrow} , u_{in\downarrow} , v_{in\downarrow} , v_{in\uparrow} \right)$ is a four-component spinor, while the matrix $\mathbb{H}$ is defined as
\begin{eqnarray}
\label{eq.hammatrix} \mathbb{H}_{ij} &=&
\left(
\begin{array}{cccc}
H_{ij\uparrow\uparrow} & H_{ij\uparrow\downarrow} & \Delta_{ij} & 0 \\ 
H_{ij\downarrow\uparrow} & H_{ij\downarrow\downarrow} & 0 & \Delta_{ij} \\
\Delta_{ij}^{\ast} & 0 & -H_{ij\downarrow\downarrow}^{\ast} & H_{ij\downarrow\uparrow}^{\ast} \\ 
0 & \Delta_{ij}^{\ast} & H_{ij\uparrow\downarrow}^{\ast} & -H_{ij\uparrow\uparrow}^{\ast}
\end{array} 
\right) ,
\end{eqnarray}
where
$H_{ij\sigma\sigma'} = \left( - t \delta_{ \langle i,j \rangle } - ( \bar{\mu} + \sigma h ) \delta_{ij} \right) \delta_{\sigma\sigma'} + H_{SO}^{\sigma\sigma'}$ and $\Delta_{ij} = \Delta \delta_{ij}$.
Here
$\bar{\mu} = \mu - V (i)$
is an {\it effective} local on-site chemical potential.
We introduce the following spin-orbit terms: $H_{SO}^{\uparrow\downarrow} = \lambda ( \delta_{ i+1,j } - \delta_{ i-1,j } )$, $H_{SO}^{\uparrow\uparrow} = H_{SO}^{\downarrow\downarrow} = 0$ and $H_{SO}^{\downarrow\uparrow} = ( H_{SO}^{\uparrow\downarrow} )^{\ast}$.

\begin{figure}[!b]
\includegraphics[width=\linewidth]{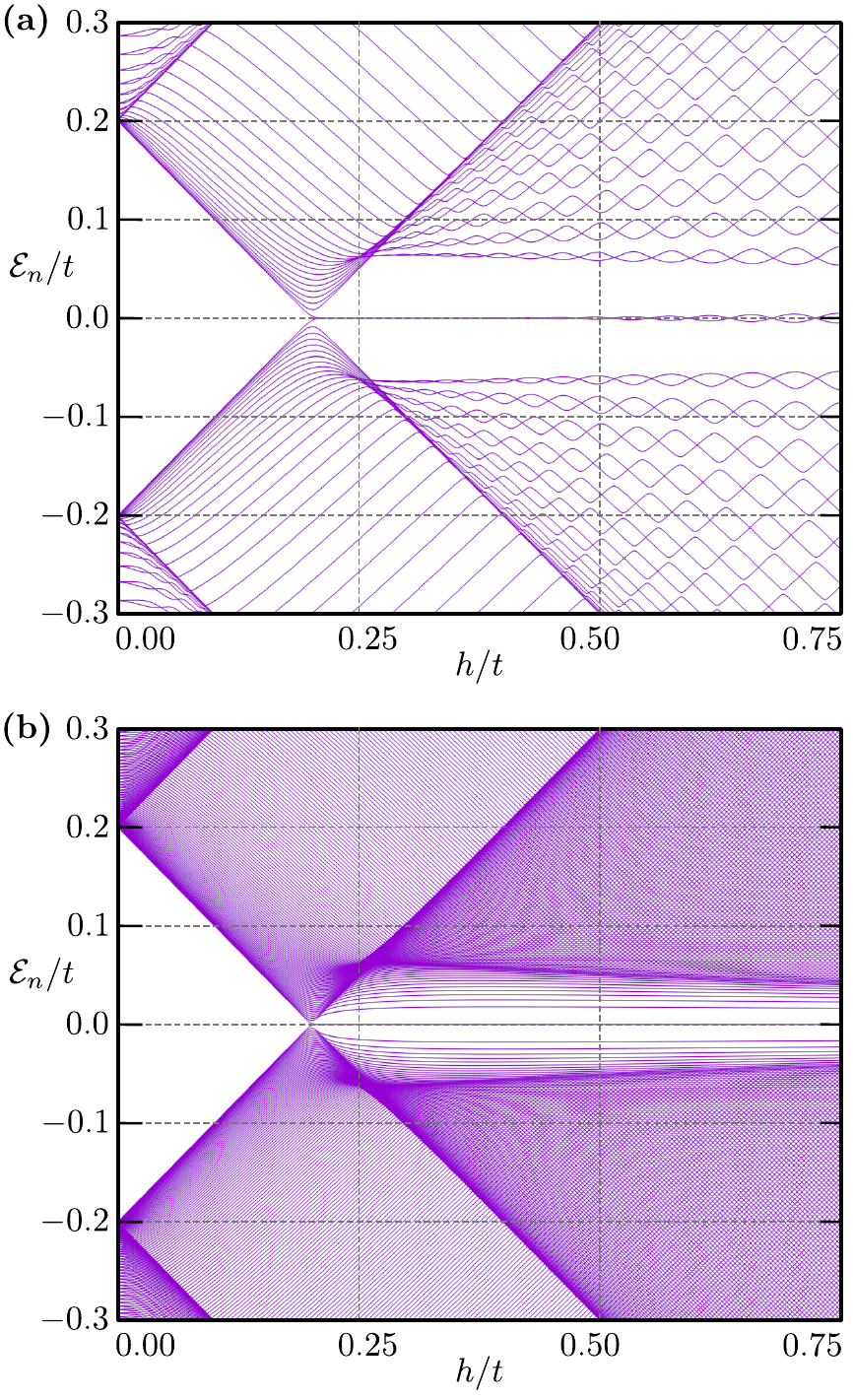}
\caption{
Eigenstates of the system for different smearing, controlled by value of $a$.
Panel (a) present results for nanowire with wall potential (large $a$), and (b) for nanowire with maximally smeared potential (small $a$).
Results obtained for $\Delta=0.2$, $\mu/t=-2$, and $\lambda/t=0.1$.
\label{fig.spec}
}
\end{figure}

\section{Numerical results}
\label{sec.num}

In our calculation, we considered the one-dimensional system with 1000 sites, while the part with homogeneous potential has $x_{\infty}-x_{0} = 100$ sites [see Eq.~(\ref{eq.pot})] -- this length of the homogeneous potential is sufficient to realize the zero-energy MBS.

Let us start by analyzing the spectrum as a function of magnetic field in case of a sharp and maximally smeared potential (Fig.~\ref{fig.spec}).
In the sharp potential case ($a \rightarrow \infty$), our system is identical with the finite nonowire [Fig.~\ref{fig.spec}(a)].
The transition form trivial to topological phase occurs at some critical magnetic field $h_{c} = \sqrt{ \Delta^{2} + \tilde{\mu}^{2} }$~\cite{sato.takahashi.09,sato.fujimoto.09,sato.takahashi.10}, where $\tilde{\mu} = \mu + 2 t$ is the chemical potential measured from the bottom of the band.
Indeed, for the discussed set of parameters, at $h = h_{c} \simeq \Delta$ the trivial gap is closed and new topological gap is reopened (see Fig.~\ref{fig.spec}).
For $h > h_{c}$, the in-gap zero energy MBS are observed, while for a relatively large $h$ we observe typical oscillation of the in-gap states around the zero-energy level.
Nevertheless, for small $h$, the MBS have zero-energies and are well localized at the boundaries of the system.

In case of the maximally smeared potential (i.e. $a \simeq 25\times 10^{-7}$), the main features of the spectrum is preserved (cf. top and bottom panel on Fig~\ref{fig.spec}).
However, additional local on-site potential lead to a situation when more sites are occupied -- this is reflected in the increased number of states observed in the presented range of energies.
Again, the zero-energy MBS are observed, while  (for $h > h_{c}$) in the initial topological gap, more in-gap states are realized [Fig~\ref{fig.spec}(b)].
This extra in-gap states are related to the states localized at the end of the system with smeared potential (and will be discussed later).

\begin{figure}[!t]
\includegraphics[width=\linewidth]{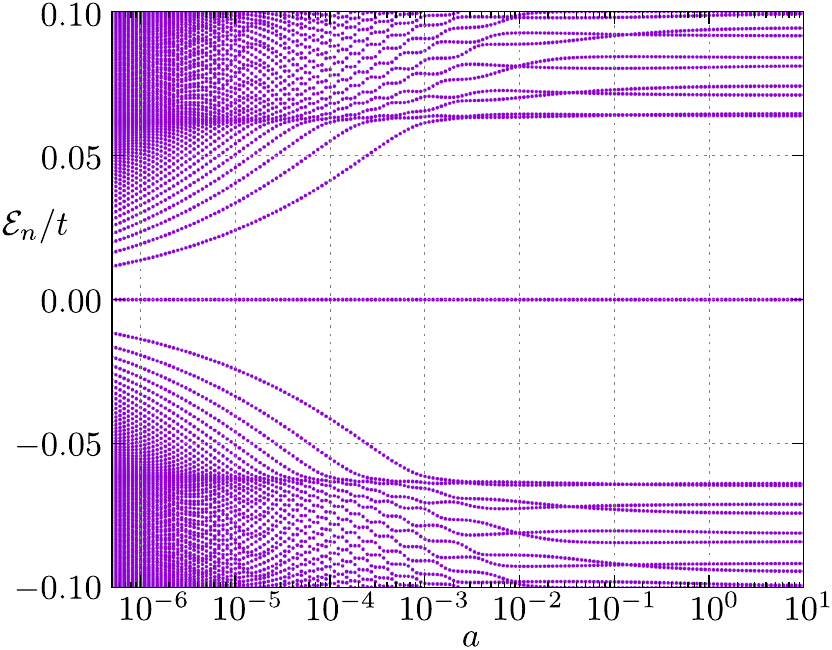}
\caption{
Evolution of the system eigenstates with smearing parameter $a$.
Results obtained for $h/t = 0.3$, $\mu/t=-2$, and $\Delta=0.2$.
\label{fig.h03}
}
\end{figure}

The above mentioned features are directly related to the potential smearing, and can be observed in Fig.~\ref{fig.h03} where we present the spectrum of the system as a function of $a$, for a fixed $h /t =0.3$.
Even for a relatively large $a \simeq 10^{-1}$ modification of the eigenvalues of the system is well visible, for states outside initial topological gap (which, for chosen sets of parameters are given in range $| \mathcal{E}_{n} | / t < 0.65$).
However, approximately around $a \simeq 10^{-3}$, the potential induce some extra states within the topological gap. 
Number of this extra in-gap states increases with decreasing $a$.
In practice, this feature can be explained as a consequence of two events.
First, the smeared potential effectively increases the number of occupied sites (for sharp potential, i.e. exactly nanowire geometry, the number of sites was $100$, while for maximally smearing potential this number increase effectively to $400$).
Second, as a consequence of the increased smearing ($a \rightarrow 0$), bigger part of the system full-fill the condition for realization of the topological phase.
Modification of the local on-site potential $\mu_{i}$ by the smearing part $V(x)$ leads to effectively decreasing the gap locally.

\begin{figure}[!b]
\includegraphics[width=\linewidth]{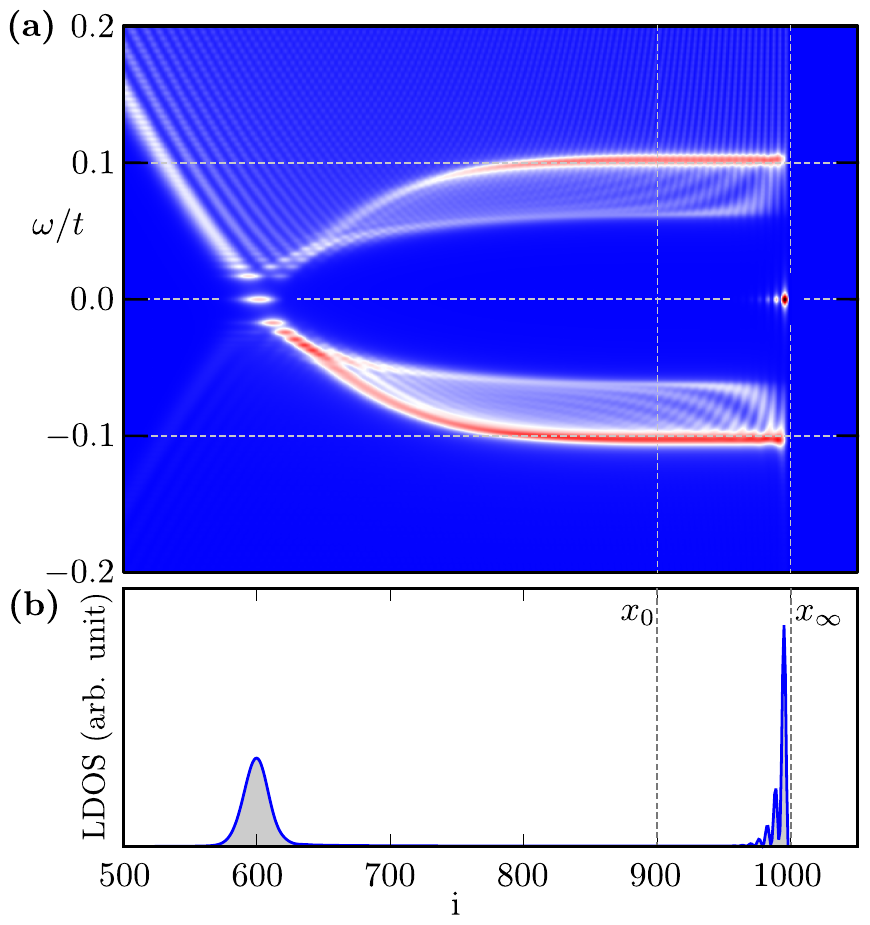}
\caption{
(a) The local density of states (LDOS) along the system and (b) their profile for $\omega/t = 0$.
Results obtained for $h/t = 0.3$, $\mu/t=-2$, $\Delta=0.2$, and $a = 10^{-6}$.
\label{fig.ldos}
}
\end{figure}

The extra in-gap states localization and the local gap suppression can be analyzed by the local density of states (LDOS), which can be found directly from solution of the BdG equations as~\cite{matsui.sato.03}:
\begin{eqnarray}
\nonumber \rho ( i , \omega ) = \sum_{n,\sigma} \left[ | u_{in\sigma} |^{2} \delta \left( \omega - \mathcal{E}_{n} \right) + | v_{in\sigma} |^{2} \delta \left( \omega + \mathcal{E}_{n} \right) \right] , \\
\label{localDOS}
\end{eqnarray}
where $\delta ( \omega )$ is the Dirac delta function, while $u_{in\sigma}$, $v_{in\sigma}$, and $\mathcal{E}_{n}$ are determined from the BdG equations. 
Results of the LDOS calculation is presented in Fig.~\ref{fig.ldos}(a), where on the right side $x_{\infty}$ we set sharp wall potential, and smearing potential is realized on left site from $x_{0}$.
Smearing of the potential lead to an effective decrease of the local gap, which is visible in form of shears-like structure in LDOS.
The occupied states are realized for $x > 600$, while the MBS are visible in form two peaks at the zero-energy.
The extra in-gap states are mostly localized on the left part of the system (along the smeared potential).
Similar situation was observed in case of the trapped fermionic superfluids~\cite{ptok.cichy.18}, where trap potential lead to an effective decrease of the gap.

Nevertheless, the gap is still open and MBS exist in the system.
However, in the presence of smearing potential from one site of the system, the localization of the MBS show different features.
In Fig.~\ref{fig.ldos}(b), we present profile of the LDOS for zero-energy states.
In case of a homogeneous system, the MBS are characterized by the oscillating wavefunctions.
Indeed, the behavior is preserved for state around $x_{\infty}$, where oscillating character is well visible.
In contrast to this, localization of the second MBS is described by a bell shaped curve (around $x = 600$). 
Additionally, the oscillating character is no longer observed.

\section{Summary}
\label{sec.sum}

Summarizing, in this paper we discussed main features of the Majorana bound states in a system with half-smeared potential.
In our investigation we assume the potential to be a quadratic function of position.
We show that the smearing potential lead to the emergence of a few extra in-gap states, and an effective decrease of the local gap.
The new in-gap states are well localized around the smearing potential.
However, changing the on-site potential by smearing modifies also the distribution of the particles in the system.
Around the last occupied sites (around the smearing potential) the effective gap decreases. 
However, the gap is still open and the Majorana states are realized in the system.
Nevertheless, the smeared potential affect the Majorana states localizations, while the states no longer have symmetric localization.
One of the Majorana states preserves their oscillating character in space, while the second one localizes in the form of a bell shaped curve.

\begin{acknowledgments}
S.B. is grateful to IT4Innovations (V\v{S}B-TU Ostrava) for hospitality during a part of the work on this project. 
This work was supported by the National Science Centre (NCN, Poland) under grants No. 
2017/25/B/ST3/02586 (S.B.) 
and
2021/43/B/ST3/02166 (A.P.). 
A.P. appreciates funding in the frame of scholarships of the Minister of Science and Higher Education (Poland) for outstanding young scientists (2019 edition, no. 818/STYP/14/2019).
\end{acknowledgments}

\bibliography{biblio}

\end{document}